\documentstyle[aps,floats,prl,twocolumn]{revtex} 
\newcommand{\bm}[1]{\mbox{\boldmath $#1$\unboldmath}}
\begin{document} 


\title{Tomonaga-Luttinger features in the
  resonant Raman spectra of quantum wires.}

\author{B.  Kramer$^{1}$, M. Sassetti$^{2}$ 
\vspace{1mm}\\
  $^{1}$I. Institut f\"ur Theoretische Physik, Universit\"at Hamburg,
  Jungiusstra\ss{}e 9, D-20355 Hamburg\\
  $^{2}$ Dipartimento di Fisica, INFM, Universit\`{a} di Genova, 
  Via Dodecaneso 33, I-16146 Genova}

\author{ {\small (September 26, 1999)}
\vspace{3mm}
}
\author{
\parbox{14cm}
{\parindent4mm
\baselineskip11pt
{\small The differential cross section for resonant Raman scattering
  from the collective modes in a one dimensional system of interacting
  electrons is calculated non-perturbatively using the bosonization
  method. The results indicate that resonant Raman spectroscopy is a
  powerful tool for studying Tomonaga-Luttinger liquid behaviour in
  quasi-one dimensional electron systems.}
\vspace{4mm}
}
}
\author{
\parbox{14cm}{
}
}
\maketitle

One dimensional (1D) electron systems are important paradigms for
studying elementary excitations. In these systems, electron-electron
correlations can be treated exactly with the bosonization technique
within the Tomonaga-Luttinger model \cite{t59,l63}. Especially, one
can rigorously show that the energetically lowest excitations are
collective \cite{h81}. The only existing modes are charge- and
spin-density excitations (CDE and SDE), with frequency-wavenumber
dispersions that are renormalized by the Coulomb repulsion and the
exchange interaction, respectively \cite{sch93,s79,v95,sch95}. In
particular, Landau-quasi particle excitations are absent in such
non-Fermi liquids, since their lifetime is vanishingly small.

One can also calculate correlation functions, say $C(\varepsilon)$,
which are experimentally observable. As a function of the variable
$\varepsilon$, typical power-law behaviors have been predicted.
Schematically,
\begin{equation}
  \label{eq:1}
  C(\varepsilon)\propto\varepsilon ^{\mu(g)}
\end{equation}
where $\mu(g)$ is in general a non-integer exponent that contains the
interaction parameter $g$. Wellknown examples are photoemission and
one-photon absorption \cite{o98}. Similar to the Fermi liquid, the
Tomonaga-Luttinger liquid appears to be of fundamental importance in
modern condensed matter theory. Therefore, directly measuring such
behavior is extremely important. Unfortunately, straightforward
experimental evidence is still missing, in spite of considerable
efforts performed on very different materials including quasi-1D
conductors and superconductors \cite{v95}. Also, predictions obtained
by mapping fractional quantum Hall states to a Luttinger liquid
\cite{muw96} have been found very difficult to confirm, as well as
Luttinger-liquid features in the dc-conductance of quantum wires
\cite{ths96}. Only recently, evidence for Luttinger behavior has been
detected in the transport properties of nano-tubes \cite{betal99}, and
in resonant tunneling through an electron island in a single-mode
quantum wire \cite{aetal99}.

A very powerful technique for studying the electronic excitations is
Raman scattering \cite{hm69,b70,p71,k75}. For energies far above the
fundamental absorption edge (off-resonance), peaks in the Raman cross
section corresponding to CDE and SDE have been identified for parallel
and perpendicular polarizations of incident and scattered light,
respectively. In resonant Raman scattering, for photon energies near
the fundamental absorption edge, polarization-insensitive structures
have been found. They have been interpreted as ``single-particle
excitations'' (``SPE'') since their dispersion corresponds roughly to
that of the pair-excitations of non-interacting electrons.

Especially in recent experiments on semiconductor quantum wires, these
polarization-insensitive features have been the subject of detailed
investigations in the regions of the intra- as well as inter-subband
transitions \cite{g91,s94,sb94,s96,pje99}. By applying the
bosonization method to the excitations in quantum wires, the physical
nature of the intra-subband ``SPE''-features has been clarified: when
approaching resonance, higher order spin density correlation functions
give rise to sharp structures in the cross-section also in parallel
polarization, with a dispersion law close to that of the SDE
\cite{sk98}.

Together with the findings at photon energies far from resonance ---
collective CDE and SDE in parallel and perpendicular polarization,
respectively --- the successful interpretation of the ``SPE''
structures suggests that Raman spectroscopy should be very promising
for testing the Tomonaga-Luttinger model for quantum wires.

In the present paper, we demonstrate that this is indeed the case. We
evaluate the differential cross-section near resonance in both
polarizations. We show that the strengths of the peaks associated with
the higher-order SDE behave according to power laws similar to
(\ref{eq:1}) when changing the photon energy and/or the temperature.
This can by no means be obtained by mean field approaches as the
random phase approximation (RPA). Confirming our predictions
experimentally, would directly indicate that quantum wires are
non-Fermi liquids.

In general, the electronic Hamiltonian of quantum wires consists of
contributions of several subbands. For describing pair excitations
with small wave numbers $q$, the subbands can be simplified to two
branches denoted by $\lambda =\pm$ with linear dispersions near the
Fermi wavenumbers $\pm k_{\rm F}$ and assumed to differ only in
``confinement energies'' $\epsilon_{j}$, measured from the minimum of
the bulk conduction band,
\begin{equation}
  \label{eq:2}
\epsilon ^{\lambda}_{j}(k)=
E_{\rm F} + \epsilon_{j} + \hbar v_{\rm F}(\lambda k - k_{\rm F})\,,   
\end{equation}
with the wave vector component $k$ in the direction of the wire. The
electron-electron interaction contains terms which couple all of the
subbands. In addition, there are matrix elements that mix only states
within a given subband. They describe backward and forward scattering
processes. While intraband forward scattering can be easily treated
within the bosonization approach \cite{ps93}, backward scattering
including the interband matrix elements, lead to severe complications,
especially near $q \approx 0$ and $T=0$ \cite{fabrizio}. However, for
describing Raman scattering, we are {\em not} interested in the
behavior at extremely small $q$. This can be used to justify a
transformation which decouples the intra- from the interband
excitations. Eventually, the Hamiltonian can be written as a quadratic
form in the corresponding charge and spin densities \cite{snk99}. In
order to demonstrate the main results of the present paper, we need
only to consider the intra-subband modes, say within the lowest
subband, $j=0$.

The bosonization technique consists of replacing the standard Fermion
fields $c_{s}^{\lambda}(k)$ associated with spin $s=\pm$ and branch
$\lambda$, by Boson fields $\Phi_{s}^{\lambda}(x,y)$. For instance,
\begin{eqnarray}
  \label{eq:3}
  c_{s}^{\lambda\dagger}(k+q)c_{s}^{\lambda}(k)&=&
  \frac{{\rm i}\lambda}{2\pi Ly}
  \int_{-\infty}^{\infty}\hspace{-3mm}{\rm d}x\, {\rm d}y \,
  {\rm e}^{{\rm i}[y(k-\lambda k_{\rm F}+q/2)+xq]}\nonumber\\
  &&\qquad\qquad\times \,\,e^{-{\rm i}\Phi_{s}^{\lambda\dagger}(x,y)}
  \,e^{-{\rm i}\Phi_{s}^{\lambda}(x,y)}\,,
\end{eqnarray}
\begin{equation}
  \label{eq:5}
  \Phi_{s}^{\lambda}(x,y)\equiv \frac{4\pi\lambda}{\sqrt{2}L}
    \sum_{q<0}\frac{{\em e}^{-{\rm i}\lambda qx}}{q}
    \sin{\left(\frac{qy}{2}\right)}
    \left[\rho^{\lambda}(\lambda q)
    +s\sigma^{\lambda}(\lambda q)\right]
\end{equation}
with $\rho^{\lambda}=\rho^{\lambda}_{+}+\rho^{\lambda}_{-}$ and
$\sigma^{\lambda}=\rho^{\lambda}_{+}-\rho^{\lambda}_{-}$ the charge
and spin densities, respectively, where
$\rho^{\lambda}_{s}(q)=\sum_{k}c_{s}^{\lambda
  \dagger}(k+q)c_{s}^{\lambda}(k)$.

This can be used to evaluate in a closed form the Fourier transform of
the correlation function
\begin{equation}
  \label{eq:6}
  \chi(q,t) = {\rm i}\Theta(t)
  \left\langle\left[N^{\dagger}(q,t),N(q,0)\right]\right\rangle
\end{equation}
which contains the generalized density operator
\begin{equation}
  \label{eq:7}
  N(q)=\sum_{k,\lambda,s}\frac{\gamma_{s}}{D(k,q)}  
  c_{s}^{\lambda\dagger}(k+q)c_{s}^{\lambda}(k)\,.
\end{equation}
The imaginary part of the former gives the differential cross section.
The quantity $\gamma_{s}$ denotes an effective optical transition
probability.  For simplicity, we assume equal transition probabilities
for parallel and perpendicular polarizations of incoming (polarization
$\bm{e}_{\rm I}$) and outgoing (polarization $\bm{e}_{\rm O}$) light,
independent of $s$,
\begin{equation}
  \label{eq:8}
  \gamma_{s}=\gamma \left(\bm{e}_{\rm I}\cdot \bm{e}_{\rm O}+
{\rm i}s|\bm{e}_{\rm I}\times \bm{e}_{\rm O}|\right)\,.  
\end{equation}
The denominator
\begin{equation}
  \label{eq:9}
  D(k,q)=E_{\rm c}(k+q) - E_{\rm v} - \hbar\omega_{\rm I} 
\end{equation}
contains the energy of incident photons $\hbar\omega_{\rm I}$, a
dispersionless valence band energy $E_{\rm v}$, and a single-subband
conduction band $E_{\rm c}(k)=\epsilon^{\lambda}_{0}(k)$ (cf.~
(\ref{eq:2})). At the first glance, this seems to be oversimplified in
view of realistic, say AlGaAs/GaAs, quantum wires.  However, it is
sufficient to explain our main results which can be straightforwardly
generalized to several subbands. It is clear from the
 (\ref{eq:3}) and (\ref{eq:5}) that (i) $N(q)$ contains all powers
of the charge and spin density operators and (ii) the cross section
can be evaluated {\em non-perturbatively}.

Out of resonance, when $\hbar\omega_{\rm I}$ is much larger than the
energy gap, $E_{\rm g} \equiv E_{\rm c}(0) - E_{\rm v}$, the energy
denominator is approximately constant. The first and second terms in
(\ref{eq:8}) give rise to peaks in the Raman spectra associated with
CDE and SDE, respectively, when inserted into (\ref{eq:7}).  This is
the ``classical selection rule''.

Closer to resonance, when the photon energy approaches $E_{\rm g}$,
higher-order correlations become important. They violate the above
selection rule. This can be seen by expanding $D(k,q)^{-1}$ in powers
of $\hbar v_{\rm F}\lambda(k-\lambda k_{\rm F})/(E_{\rm g} + E_{\rm F}
- \hbar \omega_{\rm I})$. Especially, in parallel polarization, a peak
related to a higher-order SDE has been predicted. For large photon
energies, its intensity behaves as $(E_{\rm g} + E_{\rm F} -
\hbar\omega_{\rm I})^{-4}$, in contrast to the $(E_{\rm g} + E_{\rm F}
- \hbar\omega_{\rm I})^{-2}$-behavior predicted for the SDE in
perpendicular polarization \cite{sk98}. For $\hbar \omega_{\rm I}$
very close to resonance, the non-perturbative bosonization method
leads to the characteristic non-analytic dependencies on photon energy
and temperature, as will be shown now.

In order to determine the correlation function (\ref{eq:6}) one needs
the Heisenberg operators of the charge and spin densities in the
subspace of the intraband modes of the lowest subband. For simplicity,
we assume the dispersions of the charge and spin modes to be
approximated by $\omega_{\rho}(q) = v_{\rho}(q)|q|$, with
$v_{\rho}(q)=v_{\rm F}[(1/g_{\rho}-1)\exp{(-|q|/q_{\rm int})}+1]$, and
$\omega_{\sigma}(q)=v_{\sigma}|q| $, with $v_{\sigma}=v_{\rm
  F}/g_{\sigma}$, respectively. This is justified since the
experimentally relevant region corresponds to $|q|\ll q_{\rm int}$.
The parameters $g_{\rm \rho}$ and $g_{\rm \sigma}$ describe the
strengths of Coulomb and exchange interactions, respectively.
Generally, $g_{\rm \sigma} \approx 1 > g_{\rm \rho} > 0 $ \cite{sk98}.
The cutoff $q_{\rm int}$ reflects the finite range of the repulsive
interaction in the dispersion of the CDE.

By inserting (\ref{eq:3}) into (\ref{eq:7}) and (\ref{eq:6}) one can
perform the thermal average. By taking into account translational
invariance along the wire, the cross section can be written in a
closed form as a triple integral which can computed numerically.
However, the essential physics can be extracted by the following
approximation. First, we consider contributions $\chi (q,t) \propto
\exp{({\rm i}\omega_{\sigma}(q)}t)$. These generate peak-like
structures in the Raman cross section near the frequency of the SDE.
We obtain
\begin{equation}
  \label{eq:10}
{\rm Im}\chi(q,\omega) \approx   \delta(\omega - \omega_{\sigma})  
\left[
(\bm{e}_{\rm I}\cdot \bm{e}_{\rm O})^{2}{\cal I}_{1}+
|\bm{e}_{\rm I}\times \bm{e}_{\rm O}|^{2}{\cal I}_{2}\right]  
\end{equation}
where ${\cal I}_{1}(q,\omega_{\rm I},T)$ and ${\cal
  I}_{2}(q,\omega_{\rm I},T)$ are the peaks strength in parallel and
perpendicular polarization, respectively.  Correspondingly, when
selecting $\chi (q,t) \propto \exp{({\rm i}\omega_{\rho}(q)}t)$ (since
$v_{\rho}$ approximately constant for small $q$), we get
\begin{equation}
  \label{eq:11}
{\rm Im}\chi(q,\omega) \approx \delta(\omega - \omega_{\rho})  
(\bm{e}_{\rm I}\cdot \bm{e}_{\rm O})^{2}{\cal I}_{0}\,.
\end{equation}

Equations~(\ref{eq:10}) and (\ref{eq:11}) constitute our first
general, important result: while SDE gives rise to a peak-like
structures in both polarizations, CDE appears as a peak only in
parallel and {\em not} in perpendicular configuration, even near
resonance. This can be most easily seen by considering the
lowest-order term which is $\propto \sigma \rho$ in perpendicular
polarization and this {\em cannot} give rise to a peak at the
frequency of the CDE \cite{sk98}.

Furthermore, one can prove a general theorem, namely that the terms in
a power law expansion of $N(q)$ that contribute near the frequency of
the CDE in perpendicular polarization (i) contain at least one spin
density operator, and (ii) consist always of a product of an {\em odd}
number of spin density operators multiplied by a product of charge
density operators. Terms of this kind will not produce a peak in the
corresponding cross section at the frequency of the CDE.  When
calculating the correlator, there is always a residual pair of spin
density operators, $\sigma(t)\sigma(0)$, which remains time-dependent
and destroys the coherence of the associated CDE-terms.  This
annihilates any spurious CDE-peak in the cross section.

In the following, we consider only the structures related to the SDE.
Similar results can be extracted for ${\cal I}_{0}$ Eq.(\ref{eq:11}).
The intensities of the former are (for $g_{\sigma}=1$)
\begin{equation}
  \label{eq:12}
  {\cal I}_{1}(q,\omega_{\rm I},T)=
  \frac{Lq\gamma^{2}}{12(\hbar v_{\rm F})^{2}}
  \left[\frac{q^{2}}{2}
  + \left(\frac{\pi}{\beta \hbar v_{\sigma}}\right)^{2}
  \right]\left|\frac{{\rm d}{\cal S}}{{\rm d}Q}
  \right|^{2}
\end{equation}
with $\beta ^{-1}= k_{\rm B}T$ ($k_{\rm B}$ Boltzmann constant), and
\begin{equation}
  \label{eq:13}
{\cal I}_{2}(q,\omega_{\rm I},T)=
  \frac{Lq\gamma^{2}}{(\hbar v_{\rm F})^{2}}
  \left|{\cal S}(Q,T)\right|^{2}\,.
\end{equation}
The integral
\begin{equation}
  \label{eq:14}
{\cal S}(Q,T)=\int_{0}^{\infty}{\rm d}y{\rm e}^{{\rm i}Qy}F(y)
\end{equation}
depends on the ``reduced photon wave number'' $Q=(E_{\rm g}+E_{\rm
  F} - \hbar\omega_{\rm I}+ \hbar v_{\rm F}q/2)/\hbar v_{\rm F}$. 
The function
\begin{eqnarray}
  \label{eq:15}
F(y)&=&\frac{1}{(1+q_{\rm int}^{2}y^{2})^{\mu}}
      \left[\frac{\beta \hbar v_{\sigma}}{\pi y}
        \sinh\left(\frac{\pi y}{\beta 
            \hbar v_{\sigma}}\right)\right]^{-1/2}
      \nonumber\\
&&    \nonumber\\
      &&\qquad\times\left[\frac{\beta \hbar v_{\rho}}{\pi y}
        \sinh\left(\frac{\pi y}{\beta \hbar v_{\rho}}\right)\right]
                ^{-2\mu - 1/2}
\end{eqnarray}
contains the exponent 
\begin{equation}
  \label{eq:15a}
\mu = (g_{\rho}+1/g_{\rho}-2)/8 
\end{equation}
typical for Tomonaga-Luttinger correlation functions
\cite{v95,sch95,o98}. Remarkably, it contains the parameter of the
charge interaction, though it describes SDE-related features. This
indicates that physically the higher-order SDE in parallel
configuration are ``dressed'' by CDE.

Equations~(\ref{eq:12}) to (\ref{eq:15}) constitute our second
important prediction: the dependencies of the intensities of the
SDE-peaks in resonant Raman scattering on the energy of incident
photons and/or the temperature in parallel and perpendicular
polarizations are governed by non-rational exponents that are
characteristic for the Tomonaga-Luttinger liquid and contain the
strength of the repulsive interaction between the electrons.

Let us identify in more detail the parameter regions where this
``Tomonaga-Luttinger behavior'' can be expected to be most clearly
detectable. There are three characteristic wave numbers: the inverse
of the range of the interaction $q_{\rm int}$, the wave number of the
elementary excitation $q$ and the wave number corresponding to the
temperature, $q_{\beta}=1/\beta \hbar v_{\rm F}$. We assume $q_{\rm
  int} \gg q_{\beta} > q$ since below $q_{\rm int}$ we expect the most
important interaction-induced effects. We consider interactions of
experimental relevance which correspond to $g_{\rho} > g_{0}$ with
$g_{0}$ such that $\mu (g_{0})=1/2$, i.~e. $g_{0}\approx 0.2$ and
$g_{\sigma }=1$.

For $Q > q_{\rm int}$, ${\cal I}_{n}\propto (q_{\rm int}/Q)^{4/n}$
($n=1,2$) we are still far from resonance \cite{sk98}. For $q_{\rm
  int} > Q$ we are near resonance. As long as $Q > q_{\beta}$ the
dependence on the temperature of the integral ${\cal S}(Q,T)$ does not
affect the result,
\begin{equation}
  \label{eq:16}
{\cal I}_{n}\propto \left(\frac{q_{\rm  int}}{Q}\right)^{4(1/n-\mu)}\,.
\end{equation}
For $q_{\beta}> Q$ one obtains a dependence on temperature
\begin{equation}
  \label{eq:17}
{\cal I}_{n}\propto \left(\frac{q_{\rm int}\hbar v_{\rm F}}
  {k_{\rm B}T}\right)^{4(1/n-\mu)}\,.
\end{equation}
For all of interaction parameters discussed, the ratio ${\cal
  I}_{1}/{\cal I}_{2}$ behaves independent of the interaction as
$\beta^{2}$ or $Q^{-2}$, though the energy and temperature
dependencies contain the interaction parameter. For $g_{\rho} <
g_{0}$, the behavior is similar, but cannot be treated analytically.

Traditionally, inelastic light scattering of interacting electrons has
been analyzed within RPA. This seems to work well for the non-resonant
case as it gives for quantum wires similar results for the dispersion
as the present approach. In RPA, the cross-section is related to the
electronic polarizability. By expanding into a power series in terms
of the interaction, one finds that the first term, often denoted as
$\Pi_{2}(q)$, which is independent of the interaction, contains an
energy denominator $D(k,q)^{-2}$. This is the only contribution in
{\em perpendicular} polarization \cite{k75}. It gives a peak at the
frequency of the pair excitations of the non-interacting electrons,
$v_{\rm F}|q|$.

In {\em parallel} polarization, and far from resonance, $\Pi_{2}$ can
be absorbed into a geometrical series in the interaction. This yields
only one pole --- corresponding to peak in the Raman cross section ---
at the frequency of the CDE. When approaching resonance, such that the
$k$-dependence of $D(k,q)$ has to be taken into account, $\Pi_{2}$
contributes separately \cite{p71}, and produces an additional pole at
the energy of the non-interacting electron-hole pair. The
corresponding peak intensity, however, does not show any
non-analytical power-law behavior.

In the Tomonaga-Luttinger approach, the low energy excitations are
collective. There are no modes at the energies of non-interacting
electron-hole pairs. The energetically lowest excitations are SDE with
energy $\hbar v_{\sigma}|q|$.

In principle, the renormalization of excitation frequencies could be
achieved within a self-consistent perturbational approach, generalized
to include exchange interaction, {\em but} taking into account
consistently exchange self-energy and {\em in addition} exchange
vertex corrections in $\Pi_{2}$. However, in order to obtain the above
non-analytical behavior of the intensity of SDE-peaks when approaching
resonance, these corrections should include the Coulomb interaction
{\em to infinite order}, as seen in (\ref{eq:15a}). Thus, in the
perturbative language, self-energy and vertex corrections are
responsible for the non-analytic power law behaviors of the spectra
close to resonance. This does not contradict the well known result
that far from resonance the sum of the two terms exactly cancel due to
Ward identities \cite{v95,dl74}. Indeed, the latter cannot be applied
in the presence of $k$-dependent vertices.

Presently, the existence of the ``SPE'' in the experiments on quantum
wires are well established, and consistent with our above reported
findings. Unfortunately, experimental data do not include systematic
studies of the dependencies of the peak intensities on photon energy
and/or temperature. Such studies, however, should be highly desirable
since they are expected to contribute to solving a fundamental
question of modern many-body physics, namely in how far electronic
correlations beyond mean fields are important for describing {\em
  correctly} the low-energy CDE and SDE of clean quasi-1D electron
systems.

In summary, we have pointed out that resonant Raman scattering is a
powerful tool for experimentally investigating Tomonaga-Luttinger
behavior in quasi-1D electron systems. We have shown that, when
approaching resonance, SDE-induced peaks appear in both, parallel and
perpendicular polarizations of incident and scattered photon. In
contrast, the CDE cannot produce peaks in perpendicular polarization.
We have quantitatively determined the non-analytical behavior of the
intensity of the peaks in the resonant Raman spectra that are due to
SDE. The measurement of these non-analytical dependencies on photon
energy and/or temperature predicted above would be decisive for
discovering fundamental non-Fermi liquid behavior in clean quantum
wires and represents major challenges for experiment.

We acknowledge financial support by European Uni\-on via TMR, MURST
via Cofinanziamento 98, and by the Deutsche Forschungsgemeinschaft.

\end{document}